\documentclass[aps,pra,twocolumn,floatfix,longbibliography]{revtex4-2} 

\usepackage{amsmath}
\usepackage{amssymb}
\usepackage{bm}
\usepackage{graphicx}
\usepackage{float}
\usepackage{placeins}
\usepackage{braket}
\usepackage{bbold}
\usepackage[colorlinks, linkcolor=blue, citecolor=blue, urlcolor=blue, breaklinks]{hyperref}
\usepackage{xcolor}

\usepackage{mathtext}

\newcommand  {\sub} {\mathrm{sub}}

\newcommand  {\laser} {\mathrm{laser}}
\newcommand {\at} {\mathrm{at}}

\newcommand {\inphyni} {Universit\'e C\^ote d'Azur, CNRS, Institut de Physique de Nice, France}
\newcommand {\zagreb} {Institut of Physics, Bijenička cesta 46, 10000 Zagreb, Croatia}

\begin{document}

\title{Optical interpretation of linear-optics superradiance and subradiance}
\author{S. Asselie}
\author{A. Cipris}
\altaffiliation{Present address: \zagreb}
\author{W. Guerin}
\email{william.guerin@inphyni.cnrs.fr}
\affiliation{\inphyni}

\begin{abstract}
Super- and subradiance are usually described in the framework of Dicke collective states, which is an ``atomic picture'' in which the electromagnetic field only provides an effective interaction between the atoms. Here, we discuss a complementary picture, in which we describe the propagation and scattering of light in the atomic medium, which provides a complex susceptibility and scatterers. This ``optical picture'' is valid in the linear-optics regime for disordered samples and is mainly relevant at low density, when the susceptibility and scattering cross-section can be computed from simple textbook formulas. In this picture, superradiance is a dispersion effect due to a single scattering event dressed by an effective refractive index, whereas subradiance is due to multiple scattering. We present numerical and experimental data supporting our interpretation.
%
\end{abstract}

\date{\today}

\maketitle

\section{Introduction}

Superradiance, initially introduced by Dicke \cite{Dicke:1954} and extensively studied in the 1970s-1980s \cite{Feld:1980, Gross:1982, Malcuit:1987}, is a paradigmatic example of collective effects in light-atom interaction \cite{Guerin:2017a}. It corresponds to the accelerated decay of a collection of excited atoms. More recently, the concept has been extended to the single-excitation case, in which the decay of a single quantum of excitation shared by many atoms is also accelerated compared to the single-atom decay rate, even in dilute systems
\cite{Scully:2006, Svidzinsky:2008, Scully:2009}. This collective enhancement of the decay rate of the scattered light has been studied by driving cold atoms with a weak continuous excitation, what we call the linear-optics regime \cite{Goban:2015, Araujo:2016, Roof:2016, Solano:2017, Okaba:2019, Ferioli:2021b, Pennetta:2022}. The superradiant dynamics is also visible at the switch-on \cite{EspiritoSanto:2020} and in the
coherent transmission \cite{Kwong:2015, Jennewein:2018}, in which case it can be well described by the Maxwell-Bloch equations \cite{Jennewein:2018, Svidzinsky:2015}.

The counterpart of superradiance, namely subradiance, corresponding to a slowed-down decay of the excitation, has also been observed recently, both in the linear-optics regime \cite{Guerin:2016a, Ferioli:2021a, Pennetta:2022b} and beyond \cite{Cipris:2021a, Glicenstein:2022}.

In the linear-optics regime, these collective effects are usually described using a collective-mode approach based on the coupled-dipole (CD) equations, see, e.g., \cite{Javanainen:1999, Svidzinsky:2008b, Svidzinsky:2010, Bienaime:2013, Li:2013, Schilder:2016, Sutherland:2016b, Jen:2016, Bettles:2016a, Guerin:2017b, Cipris:2021b}. In spite of its computational limitation to a few thousand atoms, the CD model is extremely useful for its broad applicability, as it allows addressing many problems. However, it does not always provide an intuitive understanding of the physical mechanisms at play.

Since what is actually measured in experiments is not collective modes but the emitted light, an optical description is also useful to gain more insight. We have started to develop such a description in two preceding articles \cite{Weiss:2021,Fofanov:2021}.

In \cite{Weiss:2021}, we have shown that linear-optics superradiance can be well described by a ``linear-dispersion theory'', initially introduced in \cite{Kuraptsev:2017} and also used with success for the switch-on dynamics \cite{Guerin:2019}. In brief, it consists in describing the collective interaction of light with the sample by a single scattering event embedded in an effective dispersive medium. Therefore, superradiance appears as a dispersive effect.

In \cite{Fofanov:2021}, we addressed subradiance and showed several signatures, computed with the CD model, of multiple scattering (or ``radiation trapping'') in the late-time dynamics after the switch-off. When the system is driven off resonance, the slow decay at late time is due to the multiple scattering of near-resonant light that originates in the switch-off--induced spectral broadening of the incident field. 

Here, we go further along these interpretations by elucidating the intrinsic link between superradiance and subradiance. Indeed, it may seem surprising that the former is due to single scattering and the latter to multiple scattering, as they are often thought to come hand in hand. In this paper, we show that there is no contradiction.

For that, we perform a numerical and experimental study of the influence of the switch-off duration and profile of the driving field. In particular, we show that, with off-resonance driving, the superradiant decay rate is a non-monotonous function of the switch-off duration: it can be larger with a finite switch-off duration than with an instantaneous switch-off. We then show that this counter-intuitive result is actually governed by \emph{single-atom physics}. The true \emph{collective} enhancement of the decay rate (superradiance) occurs only for fast switch-off, corresponding to large spectral broadening of the driving field. This leads us to a new physical picture of linear-optics superradiance: What the effective medium around the scatterer does is to filter out near-resonance light, and delay its escape by multiple scattering events, such that the early decay appears faster, because atoms respond faster off resonance \cite{Wigner:1955, Smith:1960, Bourgain:2013}. Correspondingly, a slower switch-off also suppresses subradiance for off-resonant driving, because it produces less resonant light, which strongly reduces multiple scattering.

The paper is organized as follows. In the next section, we present a numerical and experimental study of the superradiant decay rate as a function of the switch-off duration. The computations are mainly based on the linear-dispersion theory, which is computationally more efficient, but we also check that the same behavior is obtained from the coupled-dipole model. It also includes experimental data (Sec.\,\ref{sec.experiment}), which validates the numerical predictions. Then in Sec.\,\ref{sec.interpretation}, we discuss the physics behind this behavior and the consequence of this finding on the understanding of linear-optics superradiance and subradiance, before concluding.

\section{Superradiant decay rate as a function of the switch-off duration}

\subsection{Linear-dispersion theory}\label{sec.theory}

The linear-dispersion (LD) theory has been presented in detail in \cite{Weiss:2021}. It can be fruitfully used for the superradiant switch-on and switch-off dynamics \cite{Guerin:2019, Weiss:2021}. In brief, the intensity $I_{\bm{k'}}(t)$ detected in the direction $\bm{k'}$ as a function of time $t$ is given by
\begin{multline}
\label{eq.Sokolov}
I_{\bm{k'}}(t) \propto \int d^3\bm{r} \rho(\bm{r}) \, \left| \int_{-\infty}^\infty d\omega E_0(\omega) e^{-i\omega t} \right. \\ \left. \times \exp\left[i\frac{b_0(\bm{r},\bm{k'})}{2}\tilde{\alpha}(\omega)\right] \, \tilde{\alpha}(\omega) \, \exp\left[i\frac{b_0(\bm{r},\bm{k})}{2}\tilde{\alpha}(\omega)\right] \right|^2.
\end{multline}
In this equation, $\rho(\bm{r})$ is the atomic density distribution, $E_0(\omega)$ is the Fourier transform of the incident field,
\begin{equation}\label{eq.alphatild}
\tilde{\alpha}(\omega) = \frac{-1}{i + 2(\omega-\omega_\at)/\Gamma_0} 
\end{equation}
is the dimensionless atomic polarizability with $\Gamma_0 = 1/\tau_\at$ the natural linewidth and $\omega_\at$ the frequency of the atomic resonance. For simplicity, we consider motionless two-level atoms and neglect the polarization degrees of freedom, but these can easily be included \cite{Weiss:2021}. The $b_0(\bm{r},\bm{k})$ terms denote the resonant optical thickness through a part of the cloud, from the position $\bm{r}$ into the direction $\bm{k'}$, and from the incident direction $\bm{k}$ to the position $\bm{r}$. These terms can be computed, more or less easily, from the geometry of the sample. Refs. \cite{Guerin:2019, Weiss:2021} contain the formula for the simple case of a spherical Gaussian density distribution.


The meaning of Eq.\,(\ref{eq.Sokolov}) is clear: Each Fourier component of the initial field propagates through the cloud until the scattering position $\bm{r}$, propagation during which it undergoes attenuation and dephasing following Beer's law. Then it is scattered at position $\bm{r}$ with some probability and associated dephasing given by the atomic polarizability (\ref{eq.alphatild}). Finally, it propagates again through the atomic cloud until it escapes the sample. The whole process acts as a linear transfer function, which applies to the frequency components of the incident field. The temporal dependence is recovered by a Fourier transform and the intensity is computed by taking the squared modulus. Then all possible scattering positions are summed up. 

This model is valid for a single scattering only, since there is only one scattering term. Moreover, it uses the standard polarizability and scattering cross-section of a bare atom, without any shift or change due to its neighbors: it is thus valid in the dilute limit only. Finally, the intensity is summed over the scattering positions, i.e. interference effects are neglected, which is valid only for off-axis scattering averaged over the disorder configurations. When all these conditions are fulfilled, the LD theory describes well the superradiant dynamics at the switch-on and off \cite{Guerin:2019, Weiss:2021}. Note that the results do not depend on the observation direction if the atomic sample is spherically symmetric, which is the case we will consider here, with a Gaussian density profile.




\subsection{Effect of a finite switch-off duration}

\begin{figure}[t]
\centering\includegraphics{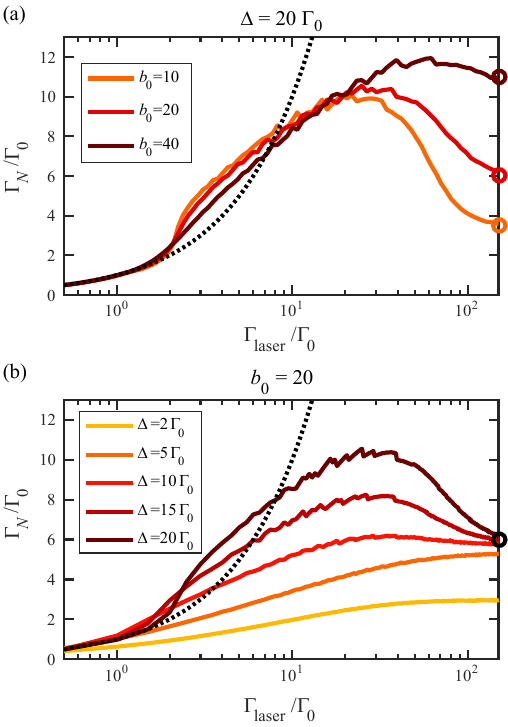}
\caption{LD prediction for the collective decay rate as a function of the laser extinction rate (solid lines, semilog scale). (a) The resonant optical thickness $b_0$ is varied with a fixed detuning $\Delta=20 \Gamma_0$. (b) The detuning $\Delta$ is varied for a fixed $b_0=20$. The dotted lines show the slow limit, when $\Gamma_N = \Gamma_\laser$, and the open circles correspond to the analytical prediction of Eq.\,(\ref{eq.instantaneous}) for an instantaneous switch-off.}
\label{fig.numerics}
\end{figure}

\begin{figure}[t]
\centering\includegraphics{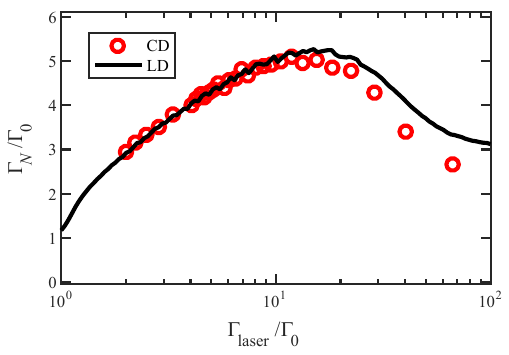}
\caption{Collective decay rate as a function of the laser extinction rate (semilog scale) computed from the coupled-dipole (CD) and linear-dispersion (LD) models, with $b_0 = 8.4$ and $\Delta = -10 \Gamma_0$.}
\label{fig.numerics_CD}
\end{figure}

We now use the LD theory in order to investigate the decay dynamics as a function of the switch-off duration of the driving field. Here, we use an exponential extinction profile with a rate $\Gamma_\mathrm{laser}$ and we fit the decay rate of the scattered light $I(t)$ with an exponential in the range $0.5 < I(t)/I_0 < 0.9$, where $I_0$ is the scattered intensity at $t=0$ when the extinction starts.

We report in Fig.\,\ref{fig.numerics} a study of the collective decay rate $\Gamma_N$ as a function of the laser extinction rate $\Gamma_\laser$. In Fig.\,\ref{fig.numerics}(a) the detuning is fixed at a large value, $\Delta=20\Gamma_0$, as this is better for achieving large superradiant decay rates \cite{Araujo:2016, Weiss:2021}, and we vary the resonant optical thickness $b_0$. In Fig.\,\ref{fig.numerics}(b) we vary the detuning for a given $b_0=20$. In the two cases, we recover the expected limiting cases. For a very small $\Gamma_\laser$ (slow limit, dotted lines in Fig.\,\ref{fig.numerics}), we get $\Gamma_N = \Gamma_\mathrm{laser}$: The decay of the scattered light follows the slow decay of the driving field. For a large $\Gamma_\laser$ (instantaneous limit, open circles in Fig.\,\ref{fig.numerics}), we recover the following analytical result \cite{Weiss:2021}, valid at large detuning:
\begin{equation}\label{eq.instantaneous}
\Gamma_N^0 = \left(1+\frac{b_0}{4}\right) \Gamma_0,
\end{equation}
where the superscript 0 denotes the instantaneous switch-off. 
The nonintuitive result reported in Fig.\,\ref{fig.numerics} is the \emph{nonmonotonous} behavior of the curves between these two limits. It shows that one can achieve a faster collective decay rate with a finite switch-off duration than with an instantaneous one. This enhancement can be large and is more pronounced for lower $b_0$ and larger detuning.

In order to check that this behavior is not an artifact of the LD theory, we have run CD simulations (in the scalar approximation with an exclusion volume, see \cite{Cipris:2021b}) with the following parameters: $b_0 = 8.4$, $\Delta = -10 \Gamma_0$, observation direction at 45$^\circ$ from the incident laser direction, averaging over the azimuthal angle and over 25 realizations. The collective decay rate is fitted in the range $0.5 < I(t)/I_0 < 0.9$ like previously. The result is shown in Fig.\,\ref{fig.numerics_CD} and confirms the behavior observed in the LD model. The small quantitative differences can be explained by the influence of subradiant modes which are included in the CD model and not in the LD one. The decay rate computed with the LD theory is thus slightly overestimated.

\subsection{Experimental data}\label{sec.experiment}

Before elaborating further on this nonintuitive effect in the next section, let us turn to its experimental verification.

\begin{figure}[b]
\centering\includegraphics{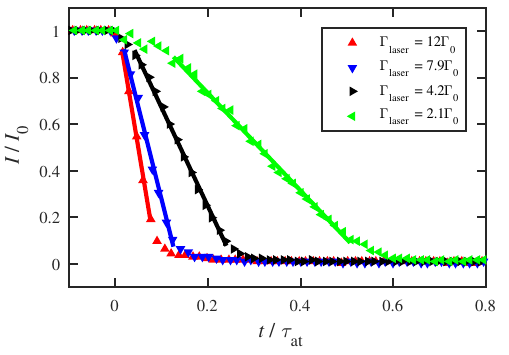}
\caption{Experimental extinction profiles of the driving laser with three different slopes and the associated linear fits.}
\label{fig.exp_profile}
\end{figure}

Our cold-atom apparatus and the way we acquire and calibrate data on the switch-off dynamics have been described in detail in previous papers \cite{Araujo:2016, Cipris:2021a}. For this new study, we adapted the switch-off method in order to make the extinction rate tunable. For that, we use a Mach-Zehnder electro-optical modulator (EOM) to control the driving beam, as in \cite{Araujo:2016}. The EOM is driven by an arbitrary function generator with which we program pulses with a tunable extinction profile. Here, we use \emph{linear} extinction profiles. For each duration, we characterize the actual switch-off profile of the driving field by measuring the extinction of the light scattered on white paper (we use the same detector as for the atomic fluorescence). We fit the extinction by a line in the range 90\% -- 10\% of the steady-state level  and define $\Gamma_\laser$ as the slope of the line. A few examples of laser extinction are reported in Fig. \ref{fig.exp_profile}. We can reach at most $\Gamma_\laser = 12 \Gamma_0$, corresponding to $\Gamma_\laser^{-1}=2.2$\,ns, limited by the bandpass of the function generator.

\begin{figure}[t]
\centering\includegraphics{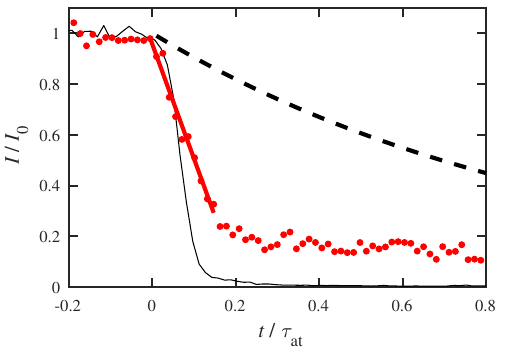}
\caption{Example of a measurement of the fluorescence decay (red dots). We also show the laser extinction (solid thin line), the fit to the data (solid thick line), and the single-atom decay for an instantaneous switch-off ($e^{-t/\tau_\at}$, dashed line) as a reference. Here, the parameters are $b_0 \simeq 22$, $\Delta = -10 \Gamma_0$, and $\Gamma_\laser \simeq 12 \Gamma_0$.}
\label{fig.example_data}
\end{figure}

We acquired several data sets for two values of the detuning $\Delta$, varying $\Gamma_\laser$. For each data set we have 12 values of the resonant optical thickness $b_0$, corresponding to different times of flight of the cold-atom cloud. For all data, the temperature is $T \approx 90 \mu$K, the saturation parameter is $s \approx 5\times 10^{-4}$ to be as close as possible to the linear-optics regime, and the density is low enough to be deeply in the dilute regime (interatomic distances much larger than the wavelength). Since the non-monotoneous behavior is better visible for small $b_0$ and large detuning, we choose $\Delta = -8\Gamma_0$ and $\Delta = -10\Gamma_0$ and $b_0$ as low as 8. Experimentally, these parameters are the most challenging as it corresponds to small amounts of scattered light. Moreover, since the driving is off resonance for the cold atoms but not for the Doppler-broadened background vapor in the vacuum chamber, a significant proportion of the detected light comes from the fluorescence of those room-temperature atoms, and, in addition, a small part from the incident laser scattered off the vacuum chamber windows. Therefore, for each data set, we also perform a full measurement without the cold atoms, which we  subtract from the measurement with the cold atoms. The resulting temporal trace is then normalized to one in the steady state (after 10 $\mu$s of driving). The superradiant decay rate is then determined from a linear fit in the range $1>I(t)/I_0>0.3$. We show an example of such a measurement (after subtraction) and fit in Fig. \ref{fig.example_data}.

\begin{figure}[t]
\centering\includegraphics{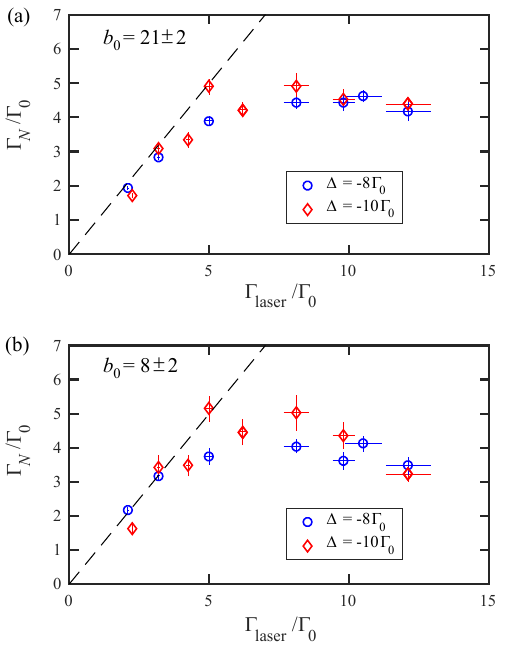}
\caption{Collective decay rate of the atomic fluorescence as a function of the laser extinction rate. Two different resonant optical depths are shown, $b_0 = 21 \pm 2$ (a) and $b_0 = 8 \pm 2$ (b). For both we show two detunings, $\Delta = -8 \Gamma_0$ (blue circles) and $\Delta = -10 \Gamma_0$ (red diamonds). The dashed line corresponds to the `slow limit' $\Gamma_N = \Gamma_\laser$. For low $b_0$ (b), we clearly observe a decrease of $\Gamma_N$ for large $\Gamma_\laser$, especially for $\Delta = -10 \Gamma_0$. The error bars are the $1\sigma$ statistical uncertainties of the fits.}
\label{fig.switchoff_exp}
\end{figure}

We report the results of our systematic study of the superradiant decay rate as a function of the laser extinction rate in Fig.\,\ref{fig.switchoff_exp} for two of the lowest $b_0$ and the two detunings. We observe well the progressive increase of $\Gamma_N$ when $\Gamma_\laser$ increases in the slow limit (left part of the curves). At higher $\Gamma_\laser$, the collective decay rate saturates and then becomes slightly smaller for the lowest $b_0$ and largest detuning, see the red diamonds in Fig.\,\ref{fig.switchoff_exp}(b). This exactly corresponds to the behavior observed in the numerical simulation. Note that since the laser switch-off profile is not exponential, the comparison with the numerical results of the previous section is only qualitative. 


\section{Interpretation}\label{sec.interpretation}

Now that the experimental results have qualitatively validated the theory, we discuss the physics behind these observations.

\subsection{Optimum decay rate}

For this, it is interesting to run more systematic numerical simulations to determine the collective decay rate as a function of the detuning and laser extinction rate. We use the LD theory for its computing efficiency. We show in Fig.\,\ref{fig.comparison_single_b0}(a) such a study for a given $b_0=20$. While all the limiting cases are well recovered (no superradiance on resonance as well as for slow extinction; $\Gamma_N^0$ for the instantaneous limit), there is a wide enhanced-decay rate region in the diagram, with a maximum approximately reached for $\Gamma_\laser \sim \Delta$. This maximum can be understood by considering the Fourier broadening associated to the pulse switch-off duration. 
Indeed, starting from the slow limit, the collective decay rate increases with $\Gamma_\laser$ up to the point when the induced broadening is such that a significant quantity of resonant light is generated, which occurs when $\Gamma_\laser \sim \Delta$.
Resonant light does not decay superradiantly \cite{Araujo:2016, Guerin:2017b} because of a high probability of doing multiple scattering, which takes time \cite{Lagendijk:1996, Labeyrie:2003, Weiss:2018}.



\subsection{Collective \textit{vs} single-atom effect}

\begin{figure}[t]
\centering\includegraphics{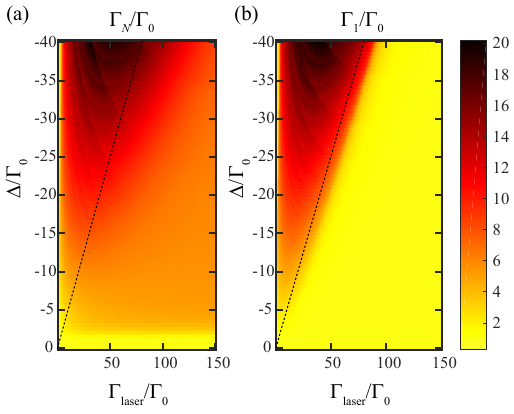}
\caption{Collective decay rate $\Gamma_N$ (a) and single-atom decay rate (b) as a function of the laser extinction rate and the detuning. For (a) the resonant optical thickness is $b_0=20$, corresponding to $\Gamma_N^0 = 6$ for an instantaneous switch-off (Eq.\,\ref{eq.instantaneous}, orange color). The dashed line is $\Delta = \Gamma_\laser/2$, which separates the single-atom physics (above, almost identical in a and b) from the collective physics (below, in a).}
\label{fig.comparison_single_b0}
\end{figure}

Another insightful fact is that we do not observe any absolute maximum of $\Gamma_N$: The maximum increases when $\Delta$ increases, independently of $b_0$. The independence with $b_0$, already visible in a large range of Fig.\,\ref{fig.numerics}(a), actually means that it is single-atom physics! To confirm this, we show in Fig.\,\ref{fig.comparison_single_b0}(b) the same computation with $b_0=0$, i.e., without the propagation parts in the LD theory. One can indeed see that all the enhanced-decay-rate region is similar, while the superradiant behavior is lost in the $\Gamma_\laser \gg \Delta$ region of the diagram. More precisely, when $\Delta \lesssim \Gamma_\laser/2$, we observe collective superradiance in Fig.\,\ref{fig.comparison_single_b0}(a), whereas for $\Delta \gtrsim \Gamma_\laser/2$, we recover single-atom physics (same in both panels). The transition between those two regimes is smooth and its exact location will depend on the extinction profile (here exponential).

This leads us to the following interpretation. In the limit of a very slow extinction, the decay of the scattered light simply is identical to the decay of the driving field. For larger $\Gamma_\laser$, the system cannot follow the driving pulse any more. One can then see the modified temporal dynamics as a \emph{distortion} of the driving pulse due to its different frequency components. The distortion effect is larger when there are near resonance frequencies because the frequency dependence of the atomic response is much stronger near resonance. 
One can thus minimize the distortion effect, and thus keep a fast fluorescence decay, by minimizing the amount of resonant light with the use of a large detuning and a not-too-fast extinction for the driving pulse. In other words, when $\Delta > \Gamma_\laser > \Gamma_0$, nothing prevents the decay dynamics of the scattered light to be faster than $\Gamma_0$, since there is little distortion.

This description works equally well with a single atom and a large sample. In the latter case, however, a fast decay is recovered even with a large spectral broadening because the effective medium around the scatterer \emph{filters out} the resonant light by the attenuation in the medium. This attenuation is not absorption but is due to a second scattering event. The corresponding light is thus delayed in time (by $1/\Gamma_0$ in average \cite{Labeyrie:2003, Weiss:2018}) and feeds the multiple-scattering orders. As a consequence, the early decay only contains single scattering from off-resonant light and thus appears faster, i.e. superradiant. The larger $b_0$, the wider the frequency filtering, and the faster the decay. 

Note that a similar filtering effect occurs at the switch-on dynamics and explains the collective behavior of the transient Rabi oscillations \cite{Guerin:2019}. Here also, a slower switch-on duration leads to less spectral broadening and, as reported in \cite{EspiritoSanto:2020}, faster damped oscillations (i.e., the steady state is reached faster).

\subsection{Subradiance}

A consequence of this interpretation is that we can predict that, if the system is driven far from resonance, a slower switch-off should suppress resonant light and thus subradiance.

\begin{figure}[t]
\centering\includegraphics{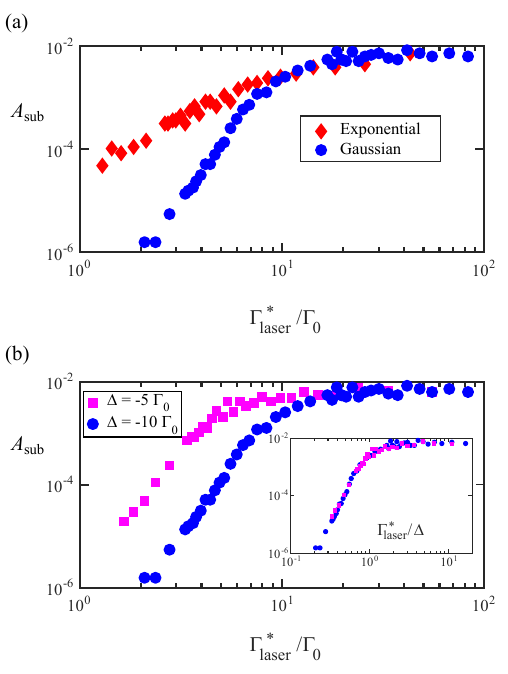}
\caption{Influence of the switch-off profile and duration of the driving field on the relative amplitude of subradiance. (a) Comparison between an exponential and a Gaussian switch-off with $b_0=8.4$ and $\Delta = -10 \Gamma_0$. $\Gamma_\laser^*$ is the width of the spectral broadening corresponding to an exponential or a Gaussian pulse \cite{footnote_broadening}. (b) Comparison between two different detunings with $b_0=8.4$ and a Gaussian switch-off profile. Inset: same data plotted as a function of $\Gamma_\laser^*/\Delta$.}
\label{fig.subradiance_switchoff}
\end{figure}

To test this prediction, we use the CD model and study the relative amplitude of the subradiant part as a function of the driving field extinction. Like previously, we use the scalar approximation with an exclusion volume and the resonant optical thickness is $b_0 = 8.4$. We average over 25 realizations and over the azimuthal angle and we compute the emitted light at 45$^\circ$ from the incident laser direction. The subradiant decay rate is fitted by a single exponential in the range $20<t/\tau_\at<30$ (different fitting window gives qualitatively similar results). We report in Fig.\,\ref{fig.subradiance_switchoff} only the relative amplitude $A_\sub$ of the subradiant decay, the lifetime being not affected.

In Fig.\,\ref{fig.subradiance_switchoff}(a) we compare two different extinction profiles with a given large detuning ($\Delta=-10\Gamma_0$): an exponential switch-off (inducing a heavy-tailed spectrum) and a Gaussian switch-off (inducing a more compact spectrum). The subradiant amplitude $A_\sub$ is the same in the instantaneous--switch-off limit but $A_\sub$ decreases when $\Gamma_\laser$ decreases, as predicted, and it decreases much more rapidly with a Gaussian switch-off, showing the influence of the spectrum tails. Similarly, we compare two different large detunings in Fig.\,\ref{fig.subradiance_switchoff}(b) in the case of a Gaussian switch-off. Here also, starting from the instantaneous limit, $A_\sub$ decreases with $\Gamma_\laser$, but the decrease occurs earlier for a larger detuning. In fact one can easily check (see inset) that the two series of data points lie on a common curve if plotted as a function of $\Gamma_\laser^*/\Delta$, and the crossover occurs for $\Gamma_\laser^*/\Delta \sim 1$, where $\Gamma_\laser^*$ is the spectral broadening \cite{footnote_broadening}. The two comparisons of Fig. \ref{fig.subradiance_switchoff} clearly put in evidence the role of the overlap between the incident spectrum and the atomic resonance, which is lower for a larger detuning or a more compact spectral broadening. Note that in all cases the time constant is not affected (here $\simeq\!6\tau_\at$), only the relative amplitude is changed, because at late times, the decay is always dominated by resonant light \cite{Fofanov:2021}. We have also checked that, if the system is driven on resonance, the subradiant amplitude is barely affected by the switch-off duration.

These observations are perfectly consistent with a multiple-scattering interpretation of subradiance \cite{Fofanov:2021} and with our optical description of superradiance.



\ \\

\section{Conclusion}

We have presented a new optical description of linear-optics superradiance, which shows that the acceleration of the fluorescence decay is due to the filtering of near-resonant light by the effective medium around the scatterer. The corresponding light is shifted to later time by multiple scattering, such that the early decay is dominated by the off-resonant scattering of a single atom, whose dynamics is faster than the natural atomic lifetime. 

We also showed that slowing-down the switch-off profile
induces a faster decay, but in that case it is due to single-atom physics. Correspondingly, subradiance is also suppressed.
This is consistent with the interpretation given in \cite{Fofanov:2021} claiming that subradiance in dilute systems is mainly due to multiple scattering of near-resonant light created by the spectral broadening related to the pulsed excitation. It is also consistent with the idea that both effects are strongly related. 

Therefore, the present work, in combination with \cite{Weiss:2021,Fofanov:2021}, gives a consistent picture of super- and subradiance in dilute and disordered samples in the linear-optics regime, only relying on classical optical phenomena such as attenuation, dispersion, and multiple scattering. This alternative description, complementary to the equally-valid collective atomic mode picture, sheds a new light on cooperative scattering in dilute disordered samples and, to our opinion, provides a more intuitive understanding of the physical mechanisms at play.

Finally, let us add that we have also shown that a controlled smoothing of the decay of the driving pulse allows accelerating the decay rate of the emitted light and avoiding the slow incoherent part due to subradiance/multiple scattering. This idea may find application in quantum-optics protocols, for instance for fast retrieval of optically-stored information with a high fidelity.


\section*{Acknowledgements}

We thank Robin Kaiser for insightful comments and constant support.
Part of this work was performed in the framework of the European Training Network ColOpt, which is funded by the European Union (EU) Horizon 2020 program under the Marie Sklodowska-Curie action, Grant Agreement No. 721465. We also acknowledge funding from the French National Research Agency (projects PACE-IN ANR19-QUAN-003 and QuaCor ANR19-CE47-0014).


\begin{thebibliography}{46}%
\makeatletter
\providecommand \@ifxundefined [1]{%
 \@ifx{#1\undefined}
}%
\providecommand \@ifnum [1]{%
 \ifnum #1\expandafter \@firstoftwo
 \else \expandafter \@secondoftwo
 \fi
}%
\providecommand \@ifx [1]{%
 \ifx #1\expandafter \@firstoftwo
 \else \expandafter \@secondoftwo
 \fi
}%
\providecommand \natexlab [1]{#1}%
\providecommand \enquote  [1]{``#1''}%
\providecommand \bibnamefont  [1]{#1}%
\providecommand \bibfnamefont [1]{#1}%
\providecommand \citenamefont [1]{#1}%
\providecommand \href@noop [0]{\@secondoftwo}%
\providecommand \href [0]{\begingroup \@sanitize@url \@href}%
\providecommand \@href[1]{\@@startlink{#1}\@@href}%
\providecommand \@@href[1]{\endgroup#1\@@endlink}%
\providecommand \@sanitize@url [0]{\catcode `\\12\catcode `\$12\catcode
  `\&12\catcode `\#12\catcode `\^12\catcode `\_12\catcode `\%12\relax}%
\providecommand \@@startlink[1]{}%
\providecommand \@@endlink[0]{}%
\providecommand \url  [0]{\begingroup\@sanitize@url \@url }%
\providecommand \@url [1]{\endgroup\@href {#1}{\urlprefix }}%
\providecommand \urlprefix  [0]{URL }%
\providecommand \Eprint [0]{\href }%
\providecommand \doibase [0]{https://doi.org/}%
\providecommand \selectlanguage [0]{\@gobble}%
\providecommand \bibinfo  [0]{\@secondoftwo}%
\providecommand \bibfield  [0]{\@secondoftwo}%
\providecommand \translation [1]{[#1]}%
\providecommand \BibitemOpen [0]{}%
\providecommand \bibitemStop [0]{}%
\providecommand \bibitemNoStop [0]{.\EOS\space}%
\providecommand \EOS [0]{\spacefactor3000\relax}%
\providecommand \BibitemShut  [1]{\csname bibitem#1\endcsname}%
\let\auto@bib@innerbib\@empty
\bibitem [{\citenamefont {Dicke}(1954)}]{Dicke:1954}%
  \BibitemOpen
  \bibfield  {author} {\bibinfo {author} {\bibfnamefont {R.~H.}\ \bibnamefont
  {Dicke}},\ }\bibfield  {title} {\bibinfo {title} {Coherence in spontaneous
  radiation processes},\ }\href@noop {} {\bibfield  {journal} {\bibinfo
  {journal} {Phys. Rev.}\ }\textbf {\bibinfo {volume} {93}},\ \bibinfo {pages}
  {99} (\bibinfo {year} {1954})}\BibitemShut {NoStop}%
\bibitem [{\citenamefont {Feld}\ and\ \citenamefont
  {{MacGillivray}}(1980)}]{Feld:1980}%
  \BibitemOpen
  \bibfield  {author} {\bibinfo {author} {\bibfnamefont {M.~S.}\ \bibnamefont
  {Feld}}\ and\ \bibinfo {author} {\bibfnamefont {J.~C.}\ \bibnamefont
  {{MacGillivray}}},\ }\bibinfo {title} {Coherent nonlinear optics. recent
  advances}\ (\bibinfo  {publisher} {Springer},\ \bibinfo {address} {Berlin},\
  \bibinfo {year} {1980})\ Chap.\ \bibinfo {chapter} {Superradiance}, pp.\
  \bibinfo {pages} {7--57}\BibitemShut {NoStop}%
\bibitem [{\citenamefont {Gross}\ and\ \citenamefont
  {Haroche}(1982)}]{Gross:1982}%
  \BibitemOpen
  \bibfield  {author} {\bibinfo {author} {\bibfnamefont {M.}~\bibnamefont
  {Gross}}\ and\ \bibinfo {author} {\bibfnamefont {S.}~\bibnamefont
  {Haroche}},\ }\bibfield  {title} {\bibinfo {title} {Superradiance: an essay
  on the theory of collective spontaneous emission},\ }\href@noop {} {\bibfield
   {journal} {\bibinfo  {journal} {Phys. Rep.}\ }\textbf {\bibinfo {volume}
  {93}},\ \bibinfo {pages} {301} (\bibinfo {year} {1982})}\BibitemShut
  {NoStop}%
\bibitem [{\citenamefont {Malcuit}\ \emph {et~al.}(1987)\citenamefont
  {Malcuit}, \citenamefont {Maki}, \citenamefont {Simkin},\ and\ \citenamefont
  {Boyd}}]{Malcuit:1987}%
  \BibitemOpen
  \bibfield  {author} {\bibinfo {author} {\bibfnamefont {M.~S.}\ \bibnamefont
  {Malcuit}}, \bibinfo {author} {\bibfnamefont {J.~J.}\ \bibnamefont {Maki}},
  \bibinfo {author} {\bibfnamefont {D.~J.}\ \bibnamefont {Simkin}},\ and\
  \bibinfo {author} {\bibfnamefont {R.~W.}\ \bibnamefont {Boyd}},\ }\bibfield
  {title} {\bibinfo {title} {Transition from superfluorescence to amplified
  spontaneous emission},\ }\href {https://doi.org/10.1103/PhysRevLett.59.1189}
  {\bibfield  {journal} {\bibinfo  {journal} {Phys. Rev. Lett.}\ }\textbf
  {\bibinfo {volume} {59}},\ \bibinfo {pages} {1189} (\bibinfo {year}
  {1987})}\BibitemShut {NoStop}%
\bibitem [{\citenamefont {Guerin}\ \emph {et~al.}(2017)\citenamefont {Guerin},
  \citenamefont {Rouabah},\ and\ \citenamefont {Kaiser}}]{Guerin:2017a}%
  \BibitemOpen
  \bibfield  {author} {\bibinfo {author} {\bibfnamefont {W.}~\bibnamefont
  {Guerin}}, \bibinfo {author} {\bibfnamefont {M.~T.}\ \bibnamefont
  {Rouabah}},\ and\ \bibinfo {author} {\bibfnamefont {R.}~\bibnamefont
  {Kaiser}},\ }\bibfield  {title} {\bibinfo {title} {Light interacting with
  atomic ensembles: collective, cooperative and mesoscopic effects},\ }\href
  {https://doi.org/10.1080/09500340.2016.1215564} {\bibfield  {journal}
  {\bibinfo  {journal} {J. Mod. Opt.}\ }\textbf {\bibinfo {volume} {64}},\
  \bibinfo {pages} {895} (\bibinfo {year} {2017})}\BibitemShut {NoStop}%
\bibitem [{\citenamefont {Scully}\ \emph {et~al.}(2006)\citenamefont {Scully},
  \citenamefont {Fry.}, \citenamefont {{Raymond~Ooi}},\ and\ \citenamefont
  {W\'odkiewicz}}]{Scully:2006}%
  \BibitemOpen
  \bibfield  {author} {\bibinfo {author} {\bibfnamefont {M.~O.}\ \bibnamefont
  {Scully}}, \bibinfo {author} {\bibfnamefont {E.~S.}\ \bibnamefont {Fry.}},
  \bibinfo {author} {\bibfnamefont {C.~H.}\ \bibnamefont {{Raymond~Ooi}}},\
  and\ \bibinfo {author} {\bibfnamefont {K.}~\bibnamefont {W\'odkiewicz}},\
  }\bibfield  {title} {\bibinfo {title} {Directed spontaneous emission from an
  extended ensemble of {$N$} atoms: Timing is everything},\ }\href@noop {}
  {\bibfield  {journal} {\bibinfo  {journal} {Phys. Rev. Lett.}\ }\textbf
  {\bibinfo {volume} {96}},\ \bibinfo {pages} {010501} (\bibinfo {year}
  {2006})}\BibitemShut {NoStop}%
\bibitem [{\citenamefont {Svidzinsky}\ \emph {et~al.}(2008)\citenamefont
  {Svidzinsky}, \citenamefont {Chang},\ and\ \citenamefont
  {Scully}}]{Svidzinsky:2008}%
  \BibitemOpen
  \bibfield  {author} {\bibinfo {author} {\bibfnamefont {A.~A.}\ \bibnamefont
  {Svidzinsky}}, \bibinfo {author} {\bibfnamefont {J.-T.}\ \bibnamefont
  {Chang}},\ and\ \bibinfo {author} {\bibfnamefont {M.~O.}\ \bibnamefont
  {Scully}},\ }\bibfield  {title} {\bibinfo {title} {Dynamical evolution of
  correlated spontaneous emission of a single photon from a uniformly excited
  cloud of {$N$} atoms},\ }\href@noop {} {\bibfield  {journal} {\bibinfo
  {journal} {Phys. Rev. Lett.}\ }\textbf {\bibinfo {volume} {100}},\ \bibinfo
  {pages} {160504} (\bibinfo {year} {2008})}\BibitemShut {NoStop}%
\bibitem [{\citenamefont {Scully}\ and\ \citenamefont
  {Svidzinsky}(2009)}]{Scully:2009}%
  \BibitemOpen
  \bibfield  {author} {\bibinfo {author} {\bibfnamefont {M.~O.}\ \bibnamefont
  {Scully}}\ and\ \bibinfo {author} {\bibfnamefont {A.~A.}\ \bibnamefont
  {Svidzinsky}},\ }\bibfield  {title} {\bibinfo {title} {The super of
  superradiance},\ }\href@noop {} {\bibfield  {journal} {\bibinfo  {journal}
  {Science}\ }\textbf {\bibinfo {volume} {325}},\ \bibinfo {pages} {1510}
  (\bibinfo {year} {2009})}\BibitemShut {NoStop}%
\bibitem [{\citenamefont {Goban}\ \emph {et~al.}(2015)\citenamefont {Goban},
  \citenamefont {Hung}, \citenamefont {Hood}, \citenamefont {Yu}, \citenamefont
  {Muniz}, \citenamefont {Painter},\ and\ \citenamefont {Kimble}}]{Goban:2015}%
  \BibitemOpen
  \bibfield  {author} {\bibinfo {author} {\bibfnamefont {A.}~\bibnamefont
  {Goban}}, \bibinfo {author} {\bibfnamefont {C.-L.}\ \bibnamefont {Hung}},
  \bibinfo {author} {\bibfnamefont {J.~D.}\ \bibnamefont {Hood}}, \bibinfo
  {author} {\bibfnamefont {S.-P.}\ \bibnamefont {Yu}}, \bibinfo {author}
  {\bibfnamefont {J.~A.}\ \bibnamefont {Muniz}}, \bibinfo {author}
  {\bibfnamefont {O.}~\bibnamefont {Painter}},\ and\ \bibinfo {author}
  {\bibfnamefont {H.~J.}\ \bibnamefont {Kimble}},\ }\bibfield  {title}
  {\bibinfo {title} {Superradiance for atoms trapped along a photonic crystal
  waveguide},\ }\href@noop {} {\bibfield  {journal} {\bibinfo  {journal} {Phys.
  Rev. Lett.}\ }\textbf {\bibinfo {volume} {115}},\ \bibinfo {pages} {063601}
  (\bibinfo {year} {2015})}\BibitemShut {NoStop}%
\bibitem [{\citenamefont {Ara\'{u}jo}\ \emph {et~al.}(2016)\citenamefont
  {Ara\'{u}jo}, \citenamefont {Kre\v{s}i\'{c}}, \citenamefont {Kaiser},\ and\
  \citenamefont {Guerin}}]{Araujo:2016}%
  \BibitemOpen
  \bibfield  {author} {\bibinfo {author} {\bibfnamefont {M.~O.}\ \bibnamefont
  {Ara\'{u}jo}}, \bibinfo {author} {\bibfnamefont {I.}~\bibnamefont
  {Kre\v{s}i\'{c}}}, \bibinfo {author} {\bibfnamefont {R.}~\bibnamefont
  {Kaiser}},\ and\ \bibinfo {author} {\bibfnamefont {W.}~\bibnamefont
  {Guerin}},\ }\bibfield  {title} {\bibinfo {title} {Superradiance in a large
  cloud of cold atoms in the linear-optics regime},\ }\href@noop {} {\bibfield
  {journal} {\bibinfo  {journal} {Phys. Rev. Lett.}\ }\textbf {\bibinfo
  {volume} {117}},\ \bibinfo {pages} {073002} (\bibinfo {year}
  {2016})}\BibitemShut {NoStop}%
\bibitem [{\citenamefont {Roof}\ \emph {et~al.}(2016)\citenamefont {Roof},
  \citenamefont {Kemp}, \citenamefont {Havey},\ and\ \citenamefont
  {Sokolov}}]{Roof:2016}%
  \BibitemOpen
  \bibfield  {author} {\bibinfo {author} {\bibfnamefont {S.~J.}\ \bibnamefont
  {Roof}}, \bibinfo {author} {\bibfnamefont {K.~J.}\ \bibnamefont {Kemp}},
  \bibinfo {author} {\bibfnamefont {M.~D.}\ \bibnamefont {Havey}},\ and\
  \bibinfo {author} {\bibfnamefont {I.~M.}\ \bibnamefont {Sokolov}},\
  }\bibfield  {title} {\bibinfo {title} {Observation of single-photon
  superradiance and the cooperative {Lamb} shift in an extended sample of cold
  atoms},\ }\href@noop {} {\bibfield  {journal} {\bibinfo  {journal} {Phys.
  Rev. Lett.}\ }\textbf {\bibinfo {volume} {117}},\ \bibinfo {pages} {073003}
  (\bibinfo {year} {2016})}\BibitemShut {NoStop}%
\bibitem [{\citenamefont {Solano}\ \emph {et~al.}(2017)\citenamefont {Solano},
  \citenamefont {Barberis-Blostein}, \citenamefont {Fatemi}, \citenamefont
  {Orozco},\ and\ \citenamefont {Rolston}}]{Solano:2017}%
  \BibitemOpen
  \bibfield  {author} {\bibinfo {author} {\bibfnamefont {P.}~\bibnamefont
  {Solano}}, \bibinfo {author} {\bibfnamefont {P.}~\bibnamefont
  {Barberis-Blostein}}, \bibinfo {author} {\bibfnamefont {F.~K.}\ \bibnamefont
  {Fatemi}}, \bibinfo {author} {\bibfnamefont {L.~A.}\ \bibnamefont {Orozco}},\
  and\ \bibinfo {author} {\bibfnamefont {S.~L.}\ \bibnamefont {Rolston}},\
  }\bibfield  {title} {\bibinfo {title} {Super-radiance reveals infinite-range
  dipole interactions through a nanofiber},\ }\href@noop {} {\bibfield
  {journal} {\bibinfo  {journal} {Nature Comm.}\ }\textbf {\bibinfo {volume}
  {8}},\ \bibinfo {pages} {1857} (\bibinfo {year} {2017})}\BibitemShut
  {NoStop}%
\bibitem [{\citenamefont {Okaba}\ \emph {et~al.}(2019)\citenamefont {Okaba},
  \citenamefont {Yu}, \citenamefont {Vincetti}, \citenamefont {Benabid},\ and\
  \citenamefont {Katori}}]{Okaba:2019}%
  \BibitemOpen
  \bibfield  {author} {\bibinfo {author} {\bibfnamefont {S.}~\bibnamefont
  {Okaba}}, \bibinfo {author} {\bibfnamefont {D.}~\bibnamefont {Yu}}, \bibinfo
  {author} {\bibfnamefont {L.}~\bibnamefont {Vincetti}}, \bibinfo {author}
  {\bibfnamefont {F.}~\bibnamefont {Benabid}},\ and\ \bibinfo {author}
  {\bibfnamefont {H.}~\bibnamefont {Katori}},\ }\bibfield  {title} {\bibinfo
  {title} {Superradiance from lattice-confined atoms inside hollow core
  fibre},\ }\href {https://doi.org/10.1038/s42005-019-0237-2} {\bibfield
  {journal} {\bibinfo  {journal} {Commun. Phys.}\ }\textbf {\bibinfo {volume}
  {2}},\ \bibinfo {pages} {136} (\bibinfo {year} {2019})}\BibitemShut {NoStop}%
\bibitem [{\citenamefont {Ferioli}\ \emph
  {et~al.}(2021{\natexlab{a}})\citenamefont {Ferioli}, \citenamefont
  {Glicenstein}, \citenamefont {Robicheaux}, \citenamefont {Sutherland},
  \citenamefont {Browaeys},\ and\ \citenamefont
  {Ferrier-Barbut}}]{Ferioli:2021b}%
  \BibitemOpen
  \bibfield  {author} {\bibinfo {author} {\bibfnamefont {G.}~\bibnamefont
  {Ferioli}}, \bibinfo {author} {\bibfnamefont {A.}~\bibnamefont
  {Glicenstein}}, \bibinfo {author} {\bibfnamefont {F.}~\bibnamefont
  {Robicheaux}}, \bibinfo {author} {\bibfnamefont {R.}~\bibnamefont
  {Sutherland}}, \bibinfo {author} {\bibfnamefont {A.}~\bibnamefont
  {Browaeys}},\ and\ \bibinfo {author} {\bibfnamefont {I.}~\bibnamefont
  {Ferrier-Barbut}},\ }\bibfield  {title} {\bibinfo {title} {Laser-driven
  superradiant ensembles of two-level atoms near {Dicke} regime},\ }\href
  {https://doi.org/10.1103/PhysRevLett.127.243602} {\bibfield  {journal}
  {\bibinfo  {journal} {Phys. Rev. Lett.}\ }\textbf {\bibinfo {volume} {127}},\
  \bibinfo {pages} {243602} (\bibinfo {year} {2021}{\natexlab{a}})}\BibitemShut
  {NoStop}%
\bibitem [{\citenamefont {Pennetta}\ \emph
  {et~al.}(2022{\natexlab{a}})\citenamefont {Pennetta}, \citenamefont {Blaha},
  \citenamefont {Johnson}, \citenamefont {Lechner}, \citenamefont
  {Schneeweiss}, \citenamefont {Volz},\ and\ \citenamefont
  {Rauschenbeutel}}]{Pennetta:2022}%
  \BibitemOpen
  \bibfield  {author} {\bibinfo {author} {\bibfnamefont {R.}~\bibnamefont
  {Pennetta}}, \bibinfo {author} {\bibfnamefont {M.}~\bibnamefont {Blaha}},
  \bibinfo {author} {\bibfnamefont {A.}~\bibnamefont {Johnson}}, \bibinfo
  {author} {\bibfnamefont {D.}~\bibnamefont {Lechner}}, \bibinfo {author}
  {\bibfnamefont {P.}~\bibnamefont {Schneeweiss}}, \bibinfo {author}
  {\bibfnamefont {J.}~\bibnamefont {Volz}},\ and\ \bibinfo {author}
  {\bibfnamefont {A.}~\bibnamefont {Rauschenbeutel}},\ }\bibfield  {title}
  {\bibinfo {title} {Collective radiative dynamics of an ensemble of cold atoms
  coupled to an optical waveguide},\ }\href
  {https://doi.org/10.1103/PhysRevLett.128.073601} {\bibfield  {journal}
  {\bibinfo  {journal} {Phys. Rev. Lett.}\ }\textbf {\bibinfo {volume} {128}},\
  \bibinfo {pages} {073601} (\bibinfo {year} {2022}{\natexlab{a}})}\BibitemShut
  {NoStop}%
\bibitem [{\citenamefont {{do Espirito Santo}}\ \emph
  {et~al.}(2020)\citenamefont {{do Espirito Santo}}, \citenamefont {Weiss},
  \citenamefont {Cipris}, \citenamefont {Kaiser}, \citenamefont {Guerin},
  \citenamefont {Bachelard},\ and\ \citenamefont
  {Schachenmayer}}]{EspiritoSanto:2020}%
  \BibitemOpen
  \bibfield  {author} {\bibinfo {author} {\bibfnamefont {T.~S.}\ \bibnamefont
  {{do Espirito Santo}}}, \bibinfo {author} {\bibfnamefont {P.}~\bibnamefont
  {Weiss}}, \bibinfo {author} {\bibfnamefont {A.}~\bibnamefont {Cipris}},
  \bibinfo {author} {\bibfnamefont {R.}~\bibnamefont {Kaiser}}, \bibinfo
  {author} {\bibfnamefont {W.}~\bibnamefont {Guerin}}, \bibinfo {author}
  {\bibfnamefont {R.}~\bibnamefont {Bachelard}},\ and\ \bibinfo {author}
  {\bibfnamefont {J.}~\bibnamefont {Schachenmayer}},\ }\bibfield  {title}
  {\bibinfo {title} {Collective excitation dynamics of a cold-atom cloud},\
  }\href {https://doi.org/10.1103/physreva.101.013617} {\bibfield  {journal}
  {\bibinfo  {journal} {Phys. Rev. A}\ }\textbf {\bibinfo {volume} {101}},\
  \bibinfo {pages} {013617} (\bibinfo {year} {2020})}\BibitemShut {NoStop}%
\bibitem [{\citenamefont {Kwong}\ \emph {et~al.}(2015)\citenamefont {Kwong},
  \citenamefont {Yang}, , \citenamefont {Delande}, \citenamefont {Pierrat},\
  and\ \citenamefont {Wilkowski}}]{Kwong:2015}%
  \BibitemOpen
  \bibfield  {author} {\bibinfo {author} {\bibfnamefont {C.~C.}\ \bibnamefont
  {Kwong}}, \bibinfo {author} {\bibfnamefont {T.}~\bibnamefont {Yang}}, ,
  \bibinfo {author} {\bibfnamefont {D.}~\bibnamefont {Delande}}, \bibinfo
  {author} {\bibfnamefont {R.}~\bibnamefont {Pierrat}},\ and\ \bibinfo {author}
  {\bibfnamefont {D.}~\bibnamefont {Wilkowski}},\ }\bibfield  {title} {\bibinfo
  {title} {Cooperative emission of a pulse train in an optically thick
  scattering medium},\ }\href@noop {} {\bibfield  {journal} {\bibinfo
  {journal} {Phys. Rev. Lett.}\ }\textbf {\bibinfo {volume} {115}},\ \bibinfo
  {pages} {223601} (\bibinfo {year} {2015})}\BibitemShut {NoStop}%
\bibitem [{\citenamefont {Jennewein}\ \emph {et~al.}(2018)\citenamefont
  {Jennewein}, \citenamefont {Brossard}, \citenamefont {Sortais}, \citenamefont
  {Browaeys}, \citenamefont {Cheinet}, \citenamefont {Robert},\ and\
  \citenamefont {Pillet}}]{Jennewein:2018}%
  \BibitemOpen
  \bibfield  {author} {\bibinfo {author} {\bibfnamefont {S.}~\bibnamefont
  {Jennewein}}, \bibinfo {author} {\bibfnamefont {L.}~\bibnamefont {Brossard}},
  \bibinfo {author} {\bibfnamefont {Y.~R.~P.}\ \bibnamefont {Sortais}},
  \bibinfo {author} {\bibfnamefont {A.}~\bibnamefont {Browaeys}}, \bibinfo
  {author} {\bibfnamefont {P.}~\bibnamefont {Cheinet}}, \bibinfo {author}
  {\bibfnamefont {J.}~\bibnamefont {Robert}},\ and\ \bibinfo {author}
  {\bibfnamefont {P.}~\bibnamefont {Pillet}},\ }\bibfield  {title} {\bibinfo
  {title} {Coherent scattering of near-resonant light by a dense, microscopic
  cloud of cold two-level atoms: Experiment versus theory},\ }\href@noop {}
  {\bibfield  {journal} {\bibinfo  {journal} {Phys. Rev. A}\ }\textbf {\bibinfo
  {volume} {97}},\ \bibinfo {pages} {053816} (\bibinfo {year}
  {2018})}\BibitemShut {NoStop}%
\bibitem [{\citenamefont {Svidzinsky}\ \emph {et~al.}(2015)\citenamefont
  {Svidzinsky}, \citenamefont {Zhang},\ and\ \citenamefont
  {Scully}}]{Svidzinsky:2015}%
  \BibitemOpen
  \bibfield  {author} {\bibinfo {author} {\bibfnamefont {A.~A.}\ \bibnamefont
  {Svidzinsky}}, \bibinfo {author} {\bibfnamefont {X.}~\bibnamefont {Zhang}},\
  and\ \bibinfo {author} {\bibfnamefont {M.~O.}\ \bibnamefont {Scully}},\
  }\bibfield  {title} {\bibinfo {title} {Quantum versus semiclassical
  description of light interaction with atomic ensembles: Revision of the
  maxwell-bloch equations and single-photon superradiance},\ }\href
  {https://doi.org/10.1103/PhysRevA.92.013801} {\bibfield  {journal} {\bibinfo
  {journal} {Phys. Rev. A}\ }\textbf {\bibinfo {volume} {92}},\ \bibinfo
  {pages} {013801} (\bibinfo {year} {2015})}\BibitemShut {NoStop}%
\bibitem [{\citenamefont {Guerin}\ \emph {et~al.}(2016)\citenamefont {Guerin},
  \citenamefont {Ara\'ujo},\ and\ \citenamefont {Kaiser}}]{Guerin:2016a}%
  \BibitemOpen
  \bibfield  {author} {\bibinfo {author} {\bibfnamefont {W.}~\bibnamefont
  {Guerin}}, \bibinfo {author} {\bibfnamefont {M.~O.}\ \bibnamefont
  {Ara\'ujo}},\ and\ \bibinfo {author} {\bibfnamefont {R.}~\bibnamefont
  {Kaiser}},\ }\bibfield  {title} {\bibinfo {title} {Subradiance in a large
  cloud of cold atoms},\ }\href@noop {} {\bibfield  {journal} {\bibinfo
  {journal} {Phys. Rev. Lett.}\ }\textbf {\bibinfo {volume} {116}},\ \bibinfo
  {pages} {083601} (\bibinfo {year} {2016})}\BibitemShut {NoStop}%
\bibitem [{\citenamefont {Ferioli}\ \emph
  {et~al.}(2021{\natexlab{b}})\citenamefont {Ferioli}, \citenamefont
  {Glicenstein}, \citenamefont {Henriet}, \citenamefont {Ferrier-Barbut},\ and\
  \citenamefont {Browaeys}}]{Ferioli:2021a}%
  \BibitemOpen
  \bibfield  {author} {\bibinfo {author} {\bibfnamefont {G.}~\bibnamefont
  {Ferioli}}, \bibinfo {author} {\bibfnamefont {A.}~\bibnamefont
  {Glicenstein}}, \bibinfo {author} {\bibfnamefont {L.}~\bibnamefont
  {Henriet}}, \bibinfo {author} {\bibfnamefont {I.}~\bibnamefont
  {Ferrier-Barbut}},\ and\ \bibinfo {author} {\bibfnamefont {A.}~\bibnamefont
  {Browaeys}},\ }\bibfield  {title} {\bibinfo {title} {Storage and release of
  subradiant excitations in a dense atomic cloud},\ }\href
  {https://doi.org/10.1103/PhysRevX.11.021031} {\bibfield  {journal} {\bibinfo
  {journal} {Phys. Rev. X}\ }\textbf {\bibinfo {volume} {11}},\ \bibinfo
  {pages} {021031} (\bibinfo {year} {2021}{\natexlab{b}})}\BibitemShut
  {NoStop}%
\bibitem [{\citenamefont {Pennetta}\ \emph
  {et~al.}(2022{\natexlab{b}})\citenamefont {Pennetta}, \citenamefont
  {Lechner}, \citenamefont {Blaha}, \citenamefont {Rauschenbeutel},
  \citenamefont {Schneeweiss},\ and\ \citenamefont {Volz}}]{Pennetta:2022b}%
  \BibitemOpen
  \bibfield  {author} {\bibinfo {author} {\bibfnamefont {R.}~\bibnamefont
  {Pennetta}}, \bibinfo {author} {\bibfnamefont {D.}~\bibnamefont {Lechner}},
  \bibinfo {author} {\bibfnamefont {M.}~\bibnamefont {Blaha}}, \bibinfo
  {author} {\bibfnamefont {A.}~\bibnamefont {Rauschenbeutel}}, \bibinfo
  {author} {\bibfnamefont {P.}~\bibnamefont {Schneeweiss}},\ and\ \bibinfo
  {author} {\bibfnamefont {J.}~\bibnamefont {Volz}},\ }\bibfield  {title}
  {\bibinfo {title} {Observation of coherent coupling between super- and
  subradiant states of an ensemble of cold atoms collectively coupled to a
  single propagating optical mode},\ }\href
  {https://doi.org/10.1103/PhysRevLett.128.203601} {\bibfield  {journal}
  {\bibinfo  {journal} {Phys. Rev. Lett.}\ }\textbf {\bibinfo {volume} {128}},\
  \bibinfo {pages} {203601} (\bibinfo {year} {2022}{\natexlab{b}})}\BibitemShut
  {NoStop}%
\bibitem [{\citenamefont {Cipris}\ \emph
  {et~al.}(2021{\natexlab{a}})\citenamefont {Cipris}, \citenamefont {Moreira},
  \citenamefont {{do~Espirito~Santo}}, \citenamefont {Weiss}, \citenamefont
  {Villas-Boas}, \citenamefont {Kaiser}, \citenamefont {Guerin},\ and\
  \citenamefont {Bachelard}}]{Cipris:2021a}%
  \BibitemOpen
  \bibfield  {author} {\bibinfo {author} {\bibfnamefont {A.}~\bibnamefont
  {Cipris}}, \bibinfo {author} {\bibfnamefont {N.~A.}\ \bibnamefont {Moreira}},
  \bibinfo {author} {\bibfnamefont {T.~S.}\ \bibnamefont
  {{do~Espirito~Santo}}}, \bibinfo {author} {\bibfnamefont {P.}~\bibnamefont
  {Weiss}}, \bibinfo {author} {\bibfnamefont {C.~J.}\ \bibnamefont
  {Villas-Boas}}, \bibinfo {author} {\bibfnamefont {R.}~\bibnamefont {Kaiser}},
  \bibinfo {author} {\bibfnamefont {W.}~\bibnamefont {Guerin}},\ and\ \bibinfo
  {author} {\bibfnamefont {R.}~\bibnamefont {Bachelard}},\ }\bibfield  {title}
  {\bibinfo {title} {Subradiance with saturated atoms: population enhancement
  of the long-lived states},\ }\href
  {https://doi.org/10.1103/PhysRevLett.126.103604} {\bibfield  {journal}
  {\bibinfo  {journal} {Phys. Rev. Lett.}\ }\textbf {\bibinfo {volume} {126}},\
  \bibinfo {pages} {103604} (\bibinfo {year} {2021}{\natexlab{a}})}\BibitemShut
  {NoStop}%
\bibitem [{\citenamefont {Glicenstein}\ \emph {et~al.}(2022)\citenamefont
  {Glicenstein}, \citenamefont {Ferioli}, \citenamefont {Browaeys},\ and\
  \citenamefont {Ferrier-Barbut}}]{Glicenstein:2022}%
  \BibitemOpen
  \bibfield  {author} {\bibinfo {author} {\bibfnamefont {A.}~\bibnamefont
  {Glicenstein}}, \bibinfo {author} {\bibfnamefont {G.}~\bibnamefont
  {Ferioli}}, \bibinfo {author} {\bibfnamefont {A.}~\bibnamefont {Browaeys}},\
  and\ \bibinfo {author} {\bibfnamefont {I.}~\bibnamefont {Ferrier-Barbut}},\
  }\bibfield  {title} {\bibinfo {title} {From superradiance to subradiance:
  exploring the many-body {Dicke} ladder},\ }\href
  {https://doi.org/10.1364/OL.451903} {\bibfield  {journal} {\bibinfo
  {journal} {Opt. Lett.}\ }\textbf {\bibinfo {volume} {47}},\ \bibinfo {pages}
  {1541} (\bibinfo {year} {2022})}\BibitemShut {NoStop}%
\bibitem [{\citenamefont {Javanainen}\ \emph {et~al.}(1999)\citenamefont
  {Javanainen}, \citenamefont {Ruostekoski}, \citenamefont {Vestergaard},\ and\
  \citenamefont {Francis}}]{Javanainen:1999}%
  \BibitemOpen
  \bibfield  {author} {\bibinfo {author} {\bibfnamefont {J.}~\bibnamefont
  {Javanainen}}, \bibinfo {author} {\bibfnamefont {J.}~\bibnamefont
  {Ruostekoski}}, \bibinfo {author} {\bibfnamefont {B.}~\bibnamefont
  {Vestergaard}},\ and\ \bibinfo {author} {\bibfnamefont {M.~R.}\ \bibnamefont
  {Francis}},\ }\bibfield  {title} {\bibinfo {title} {One-dimensional modeling
  of light propagation in dense and degenerate samples},\ }\href@noop {}
  {\bibfield  {journal} {\bibinfo  {journal} {Phys. Rev. A}\ }\textbf {\bibinfo
  {volume} {59}},\ \bibinfo {pages} {649 } (\bibinfo {year}
  {1999})}\BibitemShut {NoStop}%
\bibitem [{\citenamefont {Svidzinsky}\ and\ \citenamefont
  {Chang}(2008)}]{Svidzinsky:2008b}%
  \BibitemOpen
  \bibfield  {author} {\bibinfo {author} {\bibfnamefont {A.~A.}\ \bibnamefont
  {Svidzinsky}}\ and\ \bibinfo {author} {\bibfnamefont {J.}~\bibnamefont
  {Chang}},\ }\bibfield  {title} {\bibinfo {title} {Cooperative spontaneous
  emission as a many-body eigenvalue problem},\ }\href@noop {} {\bibfield
  {journal} {\bibinfo  {journal} {Phys. Rev. A}\ }\textbf {\bibinfo {volume}
  {77}},\ \bibinfo {pages} {043833} (\bibinfo {year} {2008})}\BibitemShut
  {NoStop}%
\bibitem [{\citenamefont {Svidzinsky}\ \emph {et~al.}(2010)\citenamefont
  {Svidzinsky}, \citenamefont {Chang},\ and\ \citenamefont
  {Scully}}]{Svidzinsky:2010}%
  \BibitemOpen
  \bibfield  {author} {\bibinfo {author} {\bibfnamefont {A.~A.}\ \bibnamefont
  {Svidzinsky}}, \bibinfo {author} {\bibfnamefont {J.}~\bibnamefont {Chang}},\
  and\ \bibinfo {author} {\bibfnamefont {M.~O.}\ \bibnamefont {Scully}},\
  }\bibfield  {title} {\bibinfo {title} {Cooperative spontaneous emission of
  {$N$} atoms: Many-body eigenstates, the effect of virtual {Lamb} shift
  processes, and analogy with radiation of {$N$} classical oscillators},\
  }\href@noop {} {\bibfield  {journal} {\bibinfo  {journal} {Phys. Rev. A}\
  }\textbf {\bibinfo {volume} {81}},\ \bibinfo {pages} {053821} (\bibinfo
  {year} {2010})}\BibitemShut {NoStop}%
\bibitem [{\citenamefont {Bienaim\'e}\ \emph {et~al.}(2013)\citenamefont
  {Bienaim\'e}, \citenamefont {Bachelard}, \citenamefont {Courteille},
  \citenamefont {Piovella},\ and\ \citenamefont {Kaiser}}]{Bienaime:2013}%
  \BibitemOpen
  \bibfield  {author} {\bibinfo {author} {\bibfnamefont {T.}~\bibnamefont
  {Bienaim\'e}}, \bibinfo {author} {\bibfnamefont {R.}~\bibnamefont
  {Bachelard}}, \bibinfo {author} {\bibfnamefont {P.~W.}\ \bibnamefont
  {Courteille}}, \bibinfo {author} {\bibfnamefont {N.}~\bibnamefont
  {Piovella}},\ and\ \bibinfo {author} {\bibfnamefont {R.}~\bibnamefont
  {Kaiser}},\ }\bibfield  {title} {\bibinfo {title} {Cooperativity in light
  scattering by cold atoms},\ }\href@noop {} {\bibfield  {journal} {\bibinfo
  {journal} {Fortschr. Phys.}\ }\textbf {\bibinfo {volume} {61}},\ \bibinfo
  {pages} {377} (\bibinfo {year} {2013})}\BibitemShut {NoStop}%
\bibitem [{\citenamefont {Li}\ \emph {et~al.}(2013)\citenamefont {Li},
  \citenamefont {Evers}, \citenamefont {Feng},\ and\ \citenamefont
  {Zhu}}]{Li:2013}%
  \BibitemOpen
  \bibfield  {author} {\bibinfo {author} {\bibfnamefont {Y.}~\bibnamefont
  {Li}}, \bibinfo {author} {\bibfnamefont {J.}~\bibnamefont {Evers}}, \bibinfo
  {author} {\bibfnamefont {W.}~\bibnamefont {Feng}},\ and\ \bibinfo {author}
  {\bibfnamefont {S.}~\bibnamefont {Zhu}},\ }\bibfield  {title} {\bibinfo
  {title} {Spectrum of collective spontaneous emission beyond the rotating-wave
  approximation},\ }\href@noop {} {\bibfield  {journal} {\bibinfo  {journal}
  {Phys. Rev. A}\ }\textbf {\bibinfo {volume} {87}},\ \bibinfo {pages} {053837}
  (\bibinfo {year} {2013})}\BibitemShut {NoStop}%
\bibitem [{\citenamefont {Schilder}\ \emph {et~al.}(2016)\citenamefont
  {Schilder}, \citenamefont {Sauvan}, \citenamefont {Hugonin}, \citenamefont
  {Jennewein}, \citenamefont {Sortais}, \citenamefont {Browaeys},\ and\
  \citenamefont {Greffet}}]{Schilder:2016}%
  \BibitemOpen
  \bibfield  {author} {\bibinfo {author} {\bibfnamefont {N.~J.}\ \bibnamefont
  {Schilder}}, \bibinfo {author} {\bibfnamefont {C.}~\bibnamefont {Sauvan}},
  \bibinfo {author} {\bibfnamefont {J.-P.}\ \bibnamefont {Hugonin}}, \bibinfo
  {author} {\bibfnamefont {S.}~\bibnamefont {Jennewein}}, \bibinfo {author}
  {\bibfnamefont {Y.~R.~P.}\ \bibnamefont {Sortais}}, \bibinfo {author}
  {\bibfnamefont {A.}~\bibnamefont {Browaeys}},\ and\ \bibinfo {author}
  {\bibfnamefont {J.-J.}\ \bibnamefont {Greffet}},\ }\bibfield  {title}
  {\bibinfo {title} {Role of polaritonic modes on light scattering from a dense
  cloud of atoms},\ }\href@noop {} {\bibfield  {journal} {\bibinfo  {journal}
  {Phys. Rev. A}\ }\textbf {\bibinfo {volume} {93}},\ \bibinfo {pages} {063835}
  (\bibinfo {year} {2016})}\BibitemShut {NoStop}%
\bibitem [{\citenamefont {Sutherland}\ and\ \citenamefont
  {Robicheaux}(2016)}]{Sutherland:2016b}%
  \BibitemOpen
  \bibfield  {author} {\bibinfo {author} {\bibfnamefont {R.~T.}\ \bibnamefont
  {Sutherland}}\ and\ \bibinfo {author} {\bibfnamefont {F.}~\bibnamefont
  {Robicheaux}},\ }\bibfield  {title} {\bibinfo {title} {Collective
  dipole-dipole interactions in an atomic array},\ }\href@noop {} {\bibfield
  {journal} {\bibinfo  {journal} {Phys. Rev. A}\ }\textbf {\bibinfo {volume}
  {94}},\ \bibinfo {pages} {013847} (\bibinfo {year} {2016})}\BibitemShut
  {NoStop}%
\bibitem [{\citenamefont {Jen}\ \emph {et~al.}(2016)\citenamefont {Jen},
  \citenamefont {Chang},\ and\ \citenamefont {Chen}}]{Jen:2016}%
  \BibitemOpen
  \bibfield  {author} {\bibinfo {author} {\bibfnamefont {H.~H.}\ \bibnamefont
  {Jen}}, \bibinfo {author} {\bibfnamefont {M.-S.}\ \bibnamefont {Chang}},\
  and\ \bibinfo {author} {\bibfnamefont {Y.-C.}\ \bibnamefont {Chen}},\
  }\bibfield  {title} {\bibinfo {title} {Cooperative single-photon subradiant
  states},\ }\href {https://doi.org/10.1103/PhysRevA.94.013803} {\bibfield
  {journal} {\bibinfo  {journal} {Phys. Rev. A}\ }\textbf {\bibinfo {volume}
  {94}},\ \bibinfo {pages} {013803} (\bibinfo {year} {2016})}\BibitemShut
  {NoStop}%
\bibitem [{\citenamefont {Bettles}\ \emph {et~al.}(2016)\citenamefont
  {Bettles}, \citenamefont {Gardiner},\ and\ \citenamefont
  {Adams}}]{Bettles:2016a}%
  \BibitemOpen
  \bibfield  {author} {\bibinfo {author} {\bibfnamefont {R.~J.}\ \bibnamefont
  {Bettles}}, \bibinfo {author} {\bibfnamefont {S.~A.}\ \bibnamefont
  {Gardiner}},\ and\ \bibinfo {author} {\bibfnamefont {C.~S.}\ \bibnamefont
  {Adams}},\ }\bibfield  {title} {\bibinfo {title} {Cooperative eigenmodes and
  scattering in one-dimensional atomic arrays},\ }\href@noop {} {\bibfield
  {journal} {\bibinfo  {journal} {Phys. Rev. A}\ }\textbf {\bibinfo {volume}
  {94}},\ \bibinfo {pages} {043844} (\bibinfo {year} {2016})}\BibitemShut
  {NoStop}%
\bibitem [{\citenamefont {Guerin}\ and\ \citenamefont
  {Kaiser}(2017)}]{Guerin:2017b}%
  \BibitemOpen
  \bibfield  {author} {\bibinfo {author} {\bibfnamefont {W.}~\bibnamefont
  {Guerin}}\ and\ \bibinfo {author} {\bibfnamefont {R.}~\bibnamefont
  {Kaiser}},\ }\bibfield  {title} {\bibinfo {title} {Population of collective
  modes in light scattering by many atoms},\ }\href@noop {} {\bibfield
  {journal} {\bibinfo  {journal} {Phys. Rev. A}\ }\textbf {\bibinfo {volume}
  {95}},\ \bibinfo {pages} {053865} (\bibinfo {year} {2017})}\BibitemShut
  {NoStop}%
\bibitem [{\citenamefont {Cipris}\ \emph
  {et~al.}(2021{\natexlab{b}})\citenamefont {Cipris}, \citenamefont
  {Bachelard}, \citenamefont {Kaiser},\ and\ \citenamefont
  {Guerin}}]{Cipris:2021b}%
  \BibitemOpen
  \bibfield  {author} {\bibinfo {author} {\bibfnamefont {A.}~\bibnamefont
  {Cipris}}, \bibinfo {author} {\bibfnamefont {R.}~\bibnamefont {Bachelard}},
  \bibinfo {author} {\bibfnamefont {R.}~\bibnamefont {Kaiser}},\ and\ \bibinfo
  {author} {\bibfnamefont {W.}~\bibnamefont {Guerin}},\ }\bibfield  {title}
  {\bibinfo {title} {{van der Waals} dephasing for {Dicke} subradiance in cold
  atomic clouds},\ }\href {https://doi.org/10.1103/PhysRevA.103.033714}
  {\bibfield  {journal} {\bibinfo  {journal} {Phys. Rev. A}\ }\textbf {\bibinfo
  {volume} {103}},\ \bibinfo {pages} {033714} (\bibinfo {year}
  {2021}{\natexlab{b}})}\BibitemShut {NoStop}%
\bibitem [{\citenamefont {Weiss}\ \emph {et~al.}(2021)\citenamefont {Weiss},
  \citenamefont {Cipris}, \citenamefont {Kaiser}, \citenamefont {Sokolov},\
  and\ \citenamefont {Guerin}}]{Weiss:2021}%
  \BibitemOpen
  \bibfield  {author} {\bibinfo {author} {\bibfnamefont {P.}~\bibnamefont
  {Weiss}}, \bibinfo {author} {\bibfnamefont {A.}~\bibnamefont {Cipris}},
  \bibinfo {author} {\bibfnamefont {R.}~\bibnamefont {Kaiser}}, \bibinfo
  {author} {\bibfnamefont {I.~M.}\ \bibnamefont {Sokolov}},\ and\ \bibinfo
  {author} {\bibfnamefont {W.}~\bibnamefont {Guerin}},\ }\bibfield  {title}
  {\bibinfo {title} {Superradiance as single scattering embedded in an
  effective medium},\ }\href {https://doi.org/10.1103/PhysRevA.103.023702}
  {\bibfield  {journal} {\bibinfo  {journal} {Phys. Rev. A}\ }\textbf {\bibinfo
  {volume} {103}},\ \bibinfo {pages} {023702} (\bibinfo {year}
  {2021})}\BibitemShut {NoStop}%
\bibitem [{\citenamefont {Fofanov}\ \emph {et~al.}(2021)\citenamefont
  {Fofanov}, \citenamefont {Sokolov}, \citenamefont {Kaiser},\ and\
  \citenamefont {Guerin}}]{Fofanov:2021}%
  \BibitemOpen
  \bibfield  {author} {\bibinfo {author} {\bibfnamefont {Y.~A.}\ \bibnamefont
  {Fofanov}}, \bibinfo {author} {\bibfnamefont {I.~M.}\ \bibnamefont
  {Sokolov}}, \bibinfo {author} {\bibfnamefont {R.}~\bibnamefont {Kaiser}},\
  and\ \bibinfo {author} {\bibfnamefont {W.}~\bibnamefont {Guerin}},\
  }\bibfield  {title} {\bibinfo {title} {Subradiance in dilute atomic ensembles
  excited by nonresonant radiation},\ }\href
  {https://doi.org/10.1103/PhysRevA.104.023705} {\bibfield  {journal} {\bibinfo
   {journal} {Phys. Rev. A}\ }\textbf {\bibinfo {volume} {104}},\ \bibinfo
  {pages} {023705} (\bibinfo {year} {2021})}\BibitemShut {NoStop}%
\bibitem [{\citenamefont {Kuraptsev}\ \emph {et~al.}(2017)\citenamefont
  {Kuraptsev}, \citenamefont {Sokolov},\ and\ \citenamefont
  {Havey}}]{Kuraptsev:2017}%
  \BibitemOpen
  \bibfield  {author} {\bibinfo {author} {\bibfnamefont {A.~S.}\ \bibnamefont
  {Kuraptsev}}, \bibinfo {author} {\bibfnamefont {I.}~\bibnamefont {Sokolov}},\
  and\ \bibinfo {author} {\bibfnamefont {M.~D.}\ \bibnamefont {Havey}},\
  }\bibfield  {title} {\bibinfo {title} {Angular distribution of single photon
  superradiance in a dilute and cold atomic ensemble},\ }\href@noop {}
  {\bibfield  {journal} {\bibinfo  {journal} {Phys. Rev. A}\ }\textbf {\bibinfo
  {volume} {96}},\ \bibinfo {pages} {023830} (\bibinfo {year}
  {2017})}\BibitemShut {NoStop}%
\bibitem [{\citenamefont {Guerin}\ \emph {et~al.}(2019)\citenamefont {Guerin},
  \citenamefont {{do Espirito Santo}}, \citenamefont {Weiss}, \citenamefont
  {Cipris}, \citenamefont {Schachenmayer}, \citenamefont {Kaiser},\ and\
  \citenamefont {Bachelard}}]{Guerin:2019}%
  \BibitemOpen
  \bibfield  {author} {\bibinfo {author} {\bibfnamefont {W.}~\bibnamefont
  {Guerin}}, \bibinfo {author} {\bibfnamefont {T.~S.}\ \bibnamefont {{do
  Espirito Santo}}}, \bibinfo {author} {\bibfnamefont {P.}~\bibnamefont
  {Weiss}}, \bibinfo {author} {\bibfnamefont {A.}~\bibnamefont {Cipris}},
  \bibinfo {author} {\bibfnamefont {J.}~\bibnamefont {Schachenmayer}}, \bibinfo
  {author} {\bibfnamefont {R.}~\bibnamefont {Kaiser}},\ and\ \bibinfo {author}
  {\bibfnamefont {R.}~\bibnamefont {Bachelard}},\ }\bibfield  {title} {\bibinfo
  {title} {Collective multi-mode vacuum {Rabi} splitting},\ }\href@noop {}
  {\bibfield  {journal} {\bibinfo  {journal} {Phys. Rev. Lett.}\ }\textbf
  {\bibinfo {volume} {123}},\ \bibinfo {pages} {243401} (\bibinfo {year}
  {2019})}\BibitemShut {NoStop}%
\bibitem [{\citenamefont {Wigner}(1955)}]{Wigner:1955}%
  \BibitemOpen
  \bibfield  {author} {\bibinfo {author} {\bibfnamefont {E.~P.}\ \bibnamefont
  {Wigner}},\ }\bibfield  {title} {\bibinfo {title} {Lower limit for the energy
  derivative of the scattering phase shift},\ }\href@noop {} {\bibfield
  {journal} {\bibinfo  {journal} {Phys. Rev.}\ }\textbf {\bibinfo {volume}
  {98}},\ \bibinfo {pages} {145} (\bibinfo {year} {1955})}\BibitemShut
  {NoStop}%
\bibitem [{\citenamefont {Smith}(1960)}]{Smith:1960}%
  \BibitemOpen
  \bibfield  {author} {\bibinfo {author} {\bibfnamefont {F.~T.}\ \bibnamefont
  {Smith}},\ }\bibfield  {title} {\bibinfo {title} {Lifetime matrix in
  collision theory},\ }\href@noop {} {\bibfield  {journal} {\bibinfo  {journal}
  {Phys. Rev.}\ }\textbf {\bibinfo {volume} {118}},\ \bibinfo {pages} {349}
  (\bibinfo {year} {1960})}\BibitemShut {NoStop}%
\bibitem [{\citenamefont {Bourgain}\ \emph {et~al.}(2013)\citenamefont
  {Bourgain}, \citenamefont {Pellegrino}, \citenamefont {Jennewein},
  \citenamefont {Sortais},\ and\ \citenamefont {Browaeys}}]{Bourgain:2013}%
  \BibitemOpen
  \bibfield  {author} {\bibinfo {author} {\bibfnamefont {R.}~\bibnamefont
  {Bourgain}}, \bibinfo {author} {\bibfnamefont {J.}~\bibnamefont
  {Pellegrino}}, \bibinfo {author} {\bibfnamefont {S.}~\bibnamefont
  {Jennewein}}, \bibinfo {author} {\bibfnamefont {Y.~R.~P.}\ \bibnamefont
  {Sortais}},\ and\ \bibinfo {author} {\bibfnamefont {A.}~\bibnamefont
  {Browaeys}},\ }\bibfield  {title} {\bibinfo {title} {Direct measurement of
  the {Wigner} time delay for the scattering of light by a single atom},\
  }\href {https://doi.org/10.1364/OL.38.001963} {\bibfield  {journal} {\bibinfo
   {journal} {Opt. Lett.}\ }\textbf {\bibinfo {volume} {38}},\ \bibinfo {pages}
  {1963} (\bibinfo {year} {2013})}\BibitemShut {NoStop}%
\bibitem [{\citenamefont {Lagendijk}\ and\ \citenamefont {van
  Tiggelen}(1996)}]{Lagendijk:1996}%
  \BibitemOpen
  \bibfield  {author} {\bibinfo {author} {\bibfnamefont {A.}~\bibnamefont
  {Lagendijk}}\ and\ \bibinfo {author} {\bibfnamefont {B.~A.}\ \bibnamefont
  {van Tiggelen}},\ }\bibfield  {title} {\bibinfo {title} {Resonant multiple
  scattering of light},\ }\href@noop {} {\bibfield  {journal} {\bibinfo
  {journal} {Phys. Rep.}\ }\textbf {\bibinfo {volume} {270}},\ \bibinfo {pages}
  {143} (\bibinfo {year} {1996})}\BibitemShut {NoStop}%
\bibitem [{\citenamefont {Labeyrie}\ \emph {et~al.}(2003)\citenamefont
  {Labeyrie}, \citenamefont {Vaujour}, \citenamefont {M\"uller}, \citenamefont
  {Delande}, \citenamefont {Miniatura}, \citenamefont {Wilkowski},\ and\
  \citenamefont {Kaiser}}]{Labeyrie:2003}%
  \BibitemOpen
  \bibfield  {author} {\bibinfo {author} {\bibfnamefont {G.}~\bibnamefont
  {Labeyrie}}, \bibinfo {author} {\bibfnamefont {E.}~\bibnamefont {Vaujour}},
  \bibinfo {author} {\bibfnamefont {C.~A.}\ \bibnamefont {M\"uller}}, \bibinfo
  {author} {\bibfnamefont {D.}~\bibnamefont {Delande}}, \bibinfo {author}
  {\bibfnamefont {C.}~\bibnamefont {Miniatura}}, \bibinfo {author}
  {\bibfnamefont {D.}~\bibnamefont {Wilkowski}},\ and\ \bibinfo {author}
  {\bibfnamefont {R.}~\bibnamefont {Kaiser}},\ }\bibfield  {title} {\bibinfo
  {title} {Slow diffusion of light in a cold atomic cloud},\ }\href@noop {}
  {\bibfield  {journal} {\bibinfo  {journal} {Phys. Rev. Lett.}\ }\textbf
  {\bibinfo {volume} {91}},\ \bibinfo {pages} {223904} (\bibinfo {year}
  {2003})}\BibitemShut {NoStop}%
\bibitem [{\citenamefont {Weiss}\ \emph {et~al.}(2018)\citenamefont {Weiss},
  \citenamefont {Ara\'ujo}, \citenamefont {Kaiser},\ and\ \citenamefont
  {Guerin}}]{Weiss:2018}%
  \BibitemOpen
  \bibfield  {author} {\bibinfo {author} {\bibfnamefont {P.}~\bibnamefont
  {Weiss}}, \bibinfo {author} {\bibfnamefont {M.~O.}\ \bibnamefont {Ara\'ujo}},
  \bibinfo {author} {\bibfnamefont {R.}~\bibnamefont {Kaiser}},\ and\ \bibinfo
  {author} {\bibfnamefont {W.}~\bibnamefont {Guerin}},\ }\bibfield  {title}
  {\bibinfo {title} {Subradiance and radiation trapping in cold atoms},\ }\href
  {https://doi.org/doi.org/10.1088/1367-2630/aac5d0} {\bibfield  {journal}
  {\bibinfo  {journal} {New. J. Phys.}\ }\textbf {\bibinfo {volume} {20}},\
  \bibinfo {pages} {063024} (\bibinfo {year} {2018})}\BibitemShut {NoStop}%
\bibitem [{foo()}]{footnote_broadening}%
  \BibitemOpen
  \href@noop {} {}\bibinfo {note} {The extinction profiles of the driving field
  intensity are defined as exp$(-\Gamma_\laser t)$ and exp$(-\Gamma_\laser^2
  t^2)$ and the CD equations are solved from the steady state at $t=0$. Then we
  defined $\Gamma_\laser^*$ as the full width at half maximum (FWHM) of the
  spectrum corresponding to pulses of light defined as exp$(-\Gamma_\laser
  |t|)$ and exp$(-\Gamma_\laser^2 t^2)$, which gives, respectively,
  $\Gamma_\laser^* = \sqrt{\sqrt{2}-1}\Gamma_\laser$ and $\Gamma_\laser^* =
  2\sqrt{\ln(2)}\Gamma_\laser$.}\BibitemShut {Stop}%
\end{thebibliography}

%

\end{document}